\documentclass[amsmath,amssymb,prl,hyperlink,twocolumn]{revtex4}

	\usepackage[pdftex]{graphicx}
	\usepackage{soul}
	\usepackage[colorlinks=true,citecolor=blue,linkcolor=magenta]{hyperref}
	\usepackage[usenames]{color}
	\usepackage{amsfonts}
	\usepackage{color}
	\usepackage{booktabs}
	\usepackage{multirow}
	\usepackage{float}
	
\usepackage{times}

\begin{document}

	\title{On-Demand Single Photons with High Extraction Efficiency and Near-Unity Indistinguishability from a Resonantly Driven Quantum Dot in a Micropillar}

	\author{Xing Ding,$^{1,2,3}$ Yu He,$^{1,2,3}$ Z.-C. Duan,$^{1,2,3}$ Niels Gregersen$^4$, M.-C. Chen$^{1,2,3}$, S. Unsleber$^5$, S. Maier$^5$, Christian Schneider$^5$, Martin Kamp$^5$, Sven H\"{o}fling$^{1,5,6}$, Chao-Yang Lu$^{1,2,3}$ and Jian-Wei Pan$^{1,2,3}$ \vspace{0.2cm}}
	
\affiliation{$^1$ Shanghai Branch, National Laboratory for Physical Sciences at Microscale and Department of Modern Physics, University of Science and Technology of China, Shanghai, 201315, China}
\affiliation{$^2$ CAS Center for Excellence and Synergetic Innovation Center in Quantum Information and Quantum Physics, University of Science and Technology of China, Hefei, Anhui 230026, China}
\affiliation{$^3$ CAS-Alibaba Quantum Computing Laboratory, Shanghai, 201315, China}
\affiliation{$^4$ DTU Fotonik, Department of Photonics Engineering, Technical University of Denmark, Building 343, DK-2800 Kongens Lyngby, Denmark}
\affiliation{$^5$ Technische Physik, Physikalisches Instit\"{a}t and Wilhelm Conrad R\"{o}ntgen-Center for Complex Material Systems, Universitat W\"{u}rzburg, Am Hubland, D-97074 W\"{u}zburg, Germany}
\affiliation{$^6$ SUPA, School of Physics and Astronomy, University of St. Andrews, St. Andrews KY16 9SS, United Kingdom}

\date{\vspace{0.1cm} submitted to PRL on 29/09/2015, accepted on 01/12/2015, published on 14/01/2016.}
	
	\begin{abstract}
Scalable photonic quantum technologies require on-demand single-photon sources with \emph{simultaneously} high levels of purity, indistinguishability, and efficiency.  These key features, however, have only been demonstrated separately in previous experiments. Here, by \textit{s}-shell pulsed resonant excitation of a Purcell-enhanced quantum dot-micropillar system, we deterministically generate resonance fluorescence single photons which, at $\pi$ pulse excitation, have an extraction efficiency of 66\%, single-photon purity of 99.1\%, and photon indistinguishability of 98.5\%. Such a single-photon source for the first time combines the features of high efficiency and near-perfect levels of purity and indistinguishabilty, and thus open the way to multi-photon experiments with semiconductor quantum dots.
	\end{abstract}
	

	\maketitle

Single-photon devices \cite{1.SPS} that deterministically emit one and only one photon at a time are central resources for scalable photonic quantum technologies \cite{2.PhotonicQI}. In particular, they are of considerable interest in Boson sampling \cite{3.BoSam}, an intermediate quantum computation where it is estimated that with 20-30 single photons one can demonstrate complex tasks that can be difficult for classical computers. To be useful for these applications, it is crucial that the single-photon source simultaneously possesses high efficiency, near perfect photon antibunching and indistinguishability---a long sought-after goal in quantum photonics.

Self-assembled quantum dots (QDs) \cite{5.QDs} have been shown to posses the highest quantum efficiency in all solid-state single-photon devices so far, and thus are promising as deterministic single-photon emitters. Tremendous progress has been made in the past decades in demonstrations of various single-photon sources \cite{Michler-Yuan, micropillar, pc, nanowire, SIL, two-photon, pulsedRF, Raman}, however, none has allowed a scalable extension to multi-photon experiments. The reason is that most previous experiments either relied on non-resonant excitation of a QD-microcavity that degraded the photon purity and indistinguishability, or used resonant excitation of a QD in a planar cavity that limited the extraction efficiency.

 In the non-resonant excitation experiments, single-photon generation efficiencies typically increased asymptotically with pump power, and a trade-off between efficiency and single-photon purity and/or indistinguishability were observed \cite{Michler-Yuan, micropillar, pc, nanowire, SIL, two-photon}. It has been shown that the recapture of carriers into the QD-micropillar due to non-resonant pumping can lead to a significant degradation of the single-photon purity \cite{purity}. Furthermore, the photon's indistinguishability can be reduced by non-resonant excitation that induces homogeneous broadening of the excited state \cite{Bennett} and uncontrolled emission time jitter from the non-radiative high-level to \textit{s}-shell relaxation \cite{Santori}. The detrimental effect of the time jitter can be more severe when using a QD-microcavity system where the Purcell-reduced radiative lifetime ($T_1$) of the QD, which bounds the single photon's coherence time ($T_2$) by $T_2\leqslant2T_1$, is comparable to or smaller than the time jitter \cite{Santori}.

To overcome these shortcomings, increasing efforts have been devoted to resonant excitation of QDs \cite{RF,Raman,pulsedRF,photonpairs,pillar}. By \textit{s}-shell resonant laser excitation on a single QD with picosecond laser pulses,  near-background-free resonance fluorescence (RF) has been obtained and allowed the observation of Rabi oscillations \cite{pulsedRF}. Under $\pi$ pulse excitation, single photons have been deterministically generated with a single-photon purity of 98.8(2)\% and indistinguishability of 97(2)\%. However, the planar cavity structure used in Ref.$\,$\cite{pulsedRF} has allowed only $\sim$6\% of the generated single photons collected by the first lens.

In this Letter, we report the first single-photon source that combines near perfect single-photon purity, indistinguishability and high extraction efficiency. By pulsed strict resonant excitation of a Purcell-enhanced QD-micropillar system, at $\pi$ pulse we obtain pulsed RF single photons with a count rate of 3.7 million per second on a silicon single-photon detector, which corresponds to an extraction efficiency of $\sim$66\% from GaAs. The RF photons show an antibuching of $g^2(0)=0.009(2)$ and a raw (corrected) two-photon interference visibility of 96.4(3)\% (98.5(4)\%). Our work opens the way to scalable multi-photon quantum information experiments with solid-state devices.

\begin{figure*}[tb]
    \centering
        \includegraphics[width=1\textwidth]{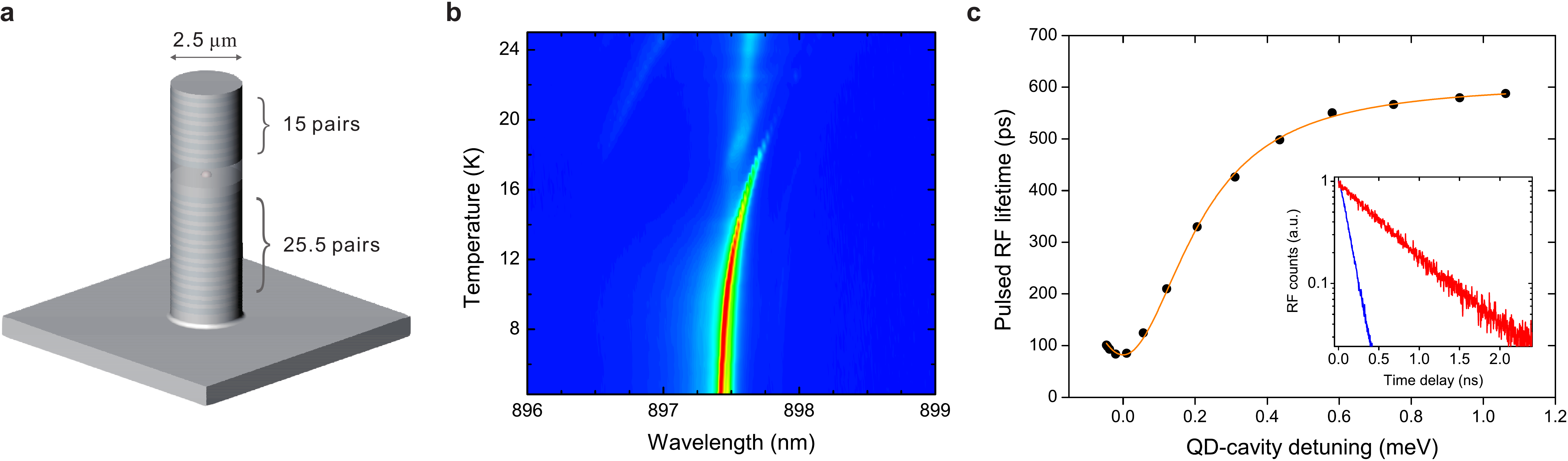}
\caption{Purcell-enhanced QD-micropillar system. (a) An illustration of a single QD embedded in a micropillar. The QD is grown via molecular beam epitaxy, embedded in a $\lambda$-thick GaAs cavity and sandwiched between 25.5 lower and 15 upper DBR stacks. Micropillars with 2.5 $\mu$m diameter were defined via electron beam lithography. (b) 2D intensity (in log scale) plot of temperature-dependent micro-photoluminescence spectra. The excitation cw laser is at 780 nm wavelength and the power is $\sim$3 nW. (c) Pulsed RF lifetime as a function of QD-cavity detuning by varying the temperature. The time-resolved data is measured using a superconducting nanowire single-photon detectors with a fast time resolution of $\sim$63 ps. The orange curve is a fit using the standard theoretical formula from ref.$\,$\cite{book}. The inset shows two examples of time-resolved RF counts at QD-cavity resonance and at far detuning.}
\label{fig:1}
\end{figure*}

\begin{figure*}[htb]
    \centering
        \includegraphics[width=0.942\textwidth]{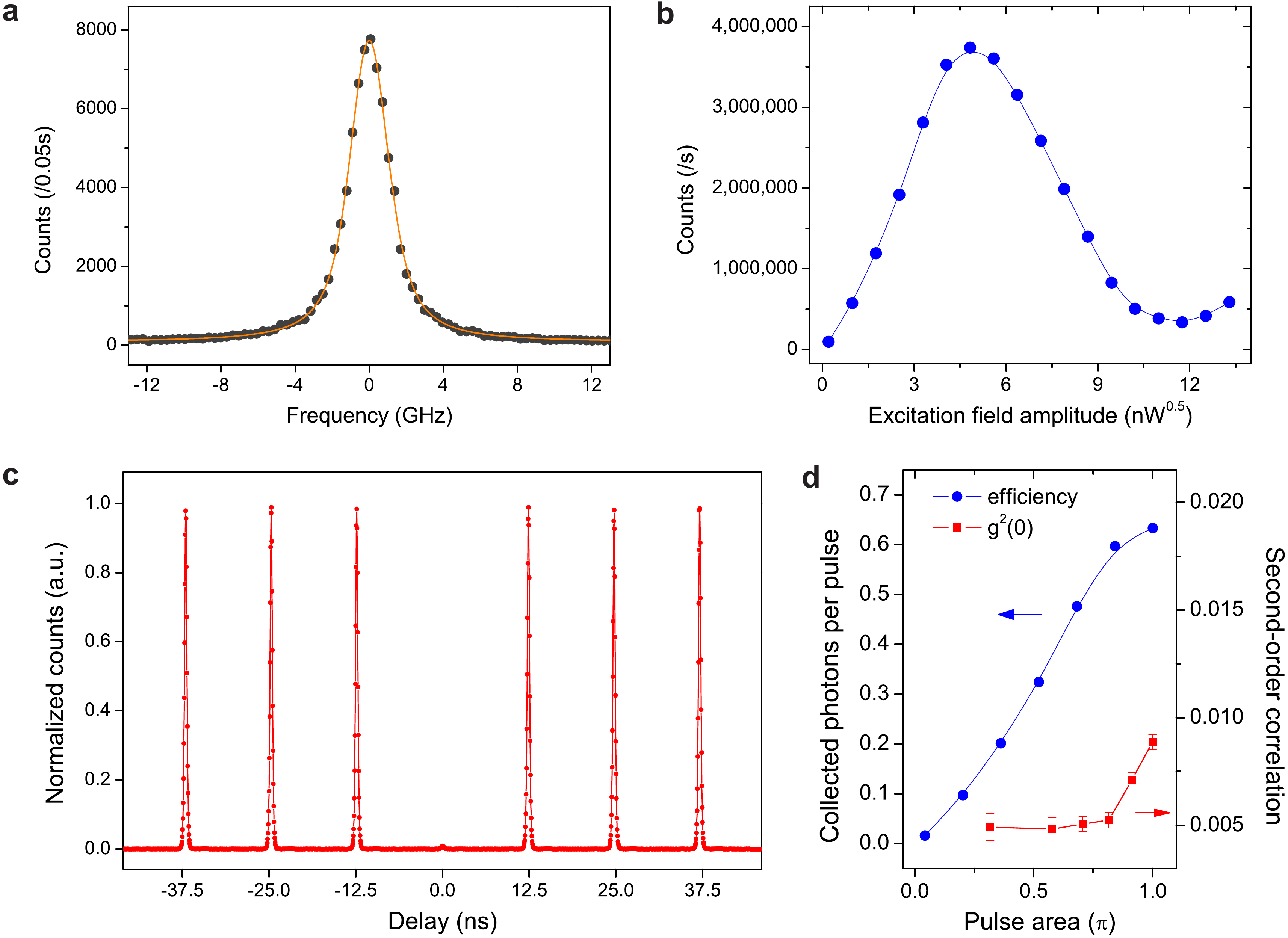}
\caption{Characterization of the pulsed RF single-photon source. (a) A high-resolution RF spectrum when excited by a $\pi$ pulse, obtained using a home-built Fabry-P\'{e}rot scanning cavity has a finesse of 170, a linewidth of 220 MHz (full width at half maximum), a free spectral range of 37.4 GHz, and a total transmission rate of 61$\%$. The orange line was fitted using a Voigt profile. (b) Detected pulsed RF counts on a silicon single-photon detector as a function of the square root of excitation laser power. The blue curve is a guide to eyes. (c) Intensity-correlation histogram of the pulsed RF photons under $\pi$ pulse excitation obtained using a Hanbury Brown and Twiss-type set-up \cite{HBT}. The second-order correlation $g^2(0)=0.009(1)$ is calculated from the integrated photon counts in the zero delay peak divided by its adjacent peak. (d) The photons collected into the first lens per pulse (generation + extraction efficiency) versus the measured single-photon purity (see text) versus pump power.}
\label{fig:2}
\end{figure*}

\begin{figure*}[tb]
    \centering
        \includegraphics[width=0.986\textwidth]{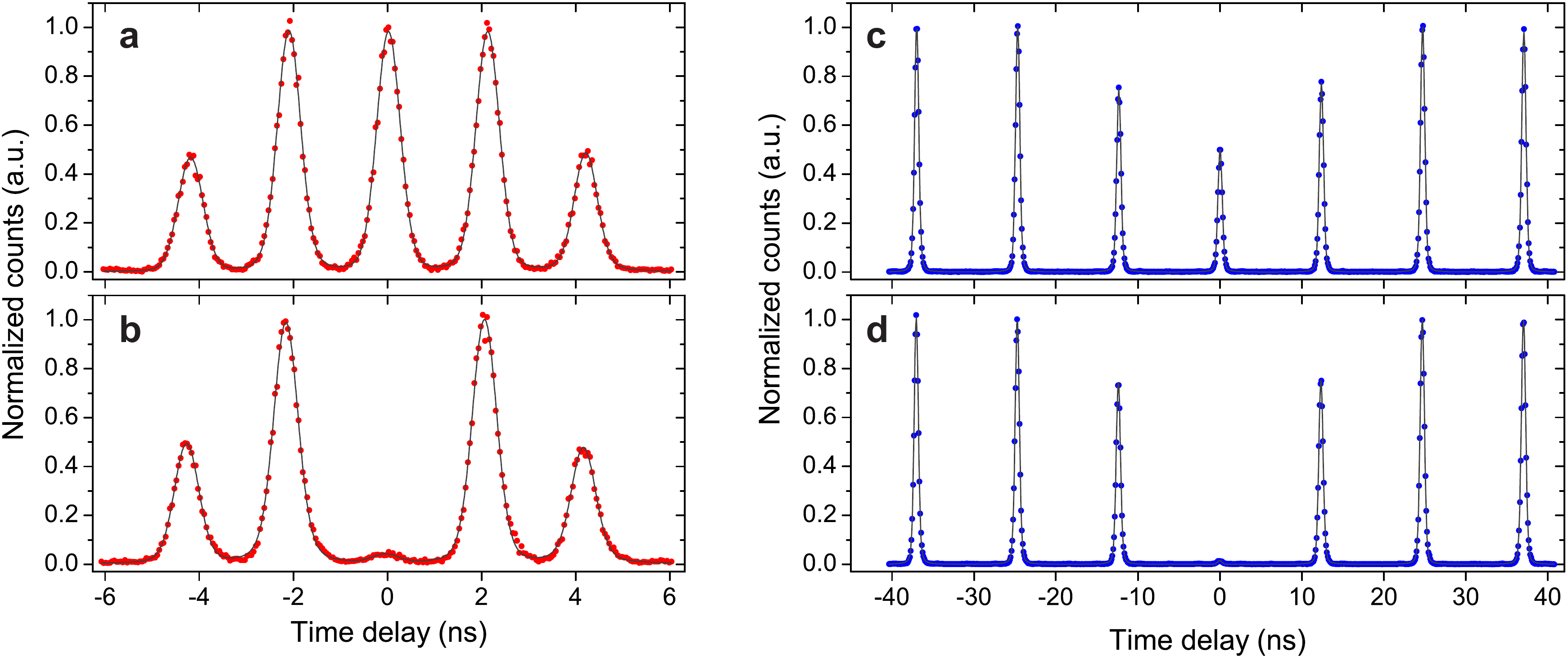}
\caption{Quantum interference between two pulsed RF single photons from a Purcell-enhanced QD-micropillar system. The time separation between the two single photons emitted from the single QD are set at 2.1 ns in \textbf{a}-\textbf{b} and 12.4 ns in \textbf{c}-\textbf{d}, respectively. The input two photons are $\pi$-pulse excited and prepared in cross (\textbf{a},\textbf{c}) and parallel (\textbf{b},\textbf{d}) polarizations, respectively. The fitting function is the convolution of exponential decay (emitter decay response) with Gaussian (photon detection time response). The area of the fitted central peaks are extracted and used to calculate the raw visibility, which is 0.964(3) and 0.959(3) for the time delay of 2.1 ns and 12.4 ns, respectively. All the data points presented are raw data without background subtraction.}
\label{fig:4}
\end{figure*}

Our experiments were performed on an InAs/GaAs self-assembled QD embedded inside a 2.5 $\mu$m diameter micropillar cavity which has 25.5 (15) $\lambda$/4-thick AlAs/GaAs mirror pairs forming the lower (upper) distributed Bragg reflectors \cite{pillar} (see Fig.$\,$1a). The device is cooled inside a cryogen-free bath cryostat with a temperature that can be finely tuned from 4.2 K to 30K. We first characterize the QD through micro-photoluminescence measurements with 780 nm laser excitation. Figure$\,$1b shows photoluminescence spectra of a single QD as a function of temperature tuning. The photoluminescence intensity reaches a plateau at a temperature range of 4.5-10 K. The quality factor of the micropillar cavity is measured to be 6124.

For pulsed resonant excitation, a Ti:Sapphire laser is used to generate laser pulses at a central wavelength of 897.44 nm and a pulse width of $\sim$3 ps. The excitation laser is further filtered with an etalon with a linewidth of 45 GHz to match the micropillar cavity. A confocal microscope is operated in a cross-polarization configuration \cite{pulsedRF}, whereby a polarizer is placed in the collection arm with its polarization perpendicular to the excitation laser suppressing the laser background by a factor exceeding $10^7$. To determine the Purcell factor of the micropillar cavity, we perform time-resolved RF measurements for different cavity-QD detunings, as plotted in Fig.$\,$1c. The shortest RF lifetime is 83.9 ps at 7.8 K (blue curve in the inset of Fig$\,$1c). At 28 K with a large ($>$1 meV) detuning, the decay time constant increases to 587.8 ps  (red curve in the inset of Fig$\,$1c). The measured lifetime as a function of detuning is well fitted by the standard weak-coupling theoretical model (orange curve) \cite{book} which allows us to deduce a Purcell factor of 6.3(4).

Figure$\,$2a shows a high-resolution spectrum of the pulsed RF, which can be best fitted using a Voigt profile with a homogeneous (Lorentzian) line width of 1.91(7) GHz and inhomogeneous (Gaussian) line width of 1.14(9) GHz. The homogeneous line width is very close to the lifetime-limited line width of 1.89 GHz. The inhomogeneous (Gaussian) component in the spectrum can be caused by spectral diffusion due to charge fluctuations in the vicinity of the QD \cite{diffusion}.

Figure$\,$2b shows the detected pulsed RF photon counts on a silicon single-photon detector as a function of incident field amplitude. We observe a Rabi oscillation which is due to coherent control of the QD two-level system \cite{Rabi}. The single-photon counts reaches the maximum for a $\pi$ pulse with an pumping laser power of 24 nW. We note that this is in stark contrast with previously experiments with non-resonant excitation where the generated single-photon counts typically grow asymptotically with excitation laser power, thus a near-unity efficiency would need very high power pumping.

At the $\pi$ pulse, we observe a count rate of 3,700,000 per second on a silicon single-photon detector, under 81 MHz repetition rate laser excitation, which gives an overall system efficiency of 4.6$\%$. The signal (RF) to background (mainly laser leakage) ratio is 40$:$1. After correcting for independently calibrated photon detection efficiency ($\sim$33$\%$), polarization extinction ($\sim$50$\%$), transmission rate in the optical path ($\sim$60$\%$, including optical window, polarizer, and two beam splitters), and single-mode fiber coupling efficiency ($\sim$72$\%$), excited-state preparation efficiency ($\sim$96$\%$) at $\pi$ pulse, and assuming perfect internal quantum efficiency of the QD, we estimate that 66$\%$ of the generated single photons are collected into the first objective lens (NA=0.68).

To further improve the signal to background ratio, we pass the RF through a 3-GHz etalon and then characterize its purity and indistinguishability. The single-photon nature of the collected RF at the $\pi$ pulse is evident from second-order coherence measurement \cite{HBT} shown in Fig.$\,$2c where nearly vanishing multi-photon probability is observed at zero delay ($g^2(0)=0.009(1)$). Figure$\,$2d summarize the combined performance of the efficiency and single-photon purity as a function of pump power. We emphasize that the high generation and extraction efficiency are obtained with little compromise of the single-photon purity, which is important for real applications in photonic quantum information processing.

Another crucial demand is that the photons should possess a high degree of indistinguishability, which is at the heart of optical quantum computing \cite{2.PhotonicQI}, Boson sampling \cite{bosam-exp} and solid-state quantum networks \cite{repeater}. Pulsed $s$-shell resonant excitation has been demonstrated to yield a near-unity indistinguishability by eliminating emission time jitter and dephasings, however, only for QDs in planar cavities. We note that the pulsed RF technique is more critically needed for QDs with large Purcell factors where the reduced radiative lifetime approaches the time jitter \cite{Santori,pulsedRF}.

The single photons' indistinguishability is tested using non-postselective two-photon Hong-Ou-Mandel interference \cite{HOM} experiments. We first adopt a similar free-space set-up as shown in Ref.$\,$\cite{pulsedRF} with a time delay of 2.1 ns between two pulses. Figure$\,$3a and 3b show time-delayed histograms of normalized two-photon counts for orthogonal and parallel polarization, respectively. An almost vanishing zero-delay peak is observed for two photons with identical polarization. In contrast, for two photons with cross polarization, the zero-delay peak has the same intensity as its adjacent peaks. We obtain a raw two-photon quantum interference visibility of 0.964(3). After correcting with the residual multiphoton probability ($g^2(0)=0.009(1)$), we obtained the corrected degrees of indistinguishability to be 0.985(4).

We further test the Hong-Ou-Mandel interference between two consecutively emitted single photons at an increased time delay of 12.4 ns---the laser pulse separation (see Ref.$\,$\cite{SIL} for a similar setup). The resulting histograms for orthogonal and parallel polarized two photons are shown in Fig.$\,$3c-d, from which we extract raw and corrected visibilities of 0.959(3) and 0.978(4), respectively. In strong contrast to Ref.$\,$\cite{SIL} that used non-resonant excitation, increasing the time delay by 10 ns has negligible effect on the photons' indistinguishability, indicating that the inhomogeneous broadening (see Fig.$\,$2a) of the pulsed RF photons are mainly from spectral diffusions at time scales much slower than 10 ns. This demonstrates the potential of this QD-micropillar device as an efficient source of a string of indistinguishable single photons, which are particularly suitable for linear optical quantum computing and Boson sampling experiments with time-bin encoding \cite{timebin}. Alternatively, one can also demultiplex the single-photon string into multiple photons at separate spatial modes.

We compare the performance of the single-photon source created in this work with heralded single photons produced by parametric down conversion \cite{SPDC} which have served as the workhorse for multi-photon interferometric experiments in the past decades \cite{2.PhotonicQI}. In the previous demonstration of the largest, eight-photon entanglement \cite{8photon}, sources of triggered single photons were generated with a count rate of 310,000/s, multi-photon emission probability of 2.9$\%$ and raw indistinguishability of 76$\%$, under a laser pump power of 880 mW. For applications such as Boson sampling \cite{bosam-exp} that needs single photons as input, the single-photon source realized here is superior, as it is ten time brighter, near perfectly indistinguishable, and requires a pump power that is 7 orders of magnitude lower.

In summary, by pulsed $s$-shell pumping a QD-micropillar system with a Purcell factor of 6.3, we have realized a high-performance single-photon source that at $\pi$ pulse excitation simultaneously achieves a generation efficiency of 96$\%$, extraction efficiency of 66$\%$, single-photon purity of 99.1$\%$ and indistinguishability of 98.6$\%$.
Such a single-photon source can be readily used to perform multi-photon interferometric experiments with a solid-state platform. Immediate applications include implementation of Boson sampling \cite{bosam-exp} with time-bin encoding using a loop-based architecture \cite{timebin}. In addition to the photonic applications, the high-efficiency extraction of transition-selective RF would also allow a fast (nanoseconds), high-fidelity single-shot readout of single electron spins \cite{readout}.

The time-jitter free, pulsed RF method is compatible with higher Purcell factor, which can allow photon extraction to be further improved in the future by optimized microcavity fabrications, without comprising the single-photon purity and indistinguishability. The current overall system efficiency---4.6$\%$, the highest reported in QDs---can also be improved using techniques such as orthogonal excitation and detection of RF \cite{RF,photonpairs}, near-unity-efficiency superconducting nanowire single-photon detection \cite{detectors}, and antireflection coatings of the optical elements.

\textit{note}: After this paper was submitted, we became aware of a related work on arXiv \cite{noteadd}.

\vspace{0.1cm}
\noindent \textit{Acknowledgement}: This work was supported by the National Natural Science Foundation of China, the Chinese Academy of Sciences, and the National Fundamental Research Program. We acknowledge financial support by the State of Bavaria and the German Ministry of Education and Research (BMBF) within the projects Q.com-H and the Chist-era project SSQN. NG acknowledges support from Danish Research Council for Technology and Production.

* email: cylu@ustc.edu.cn, pan@ustc.edu.cn

\end{document}